\documentclass[aip,jcp,amsmath,amssymb,
 reprint,%
]{revtex4-1}

\usepackage{graphicx}
\usepackage{dcolumn}
\usepackage{bm}
\usepackage{booktabs}
\usepackage{amsmath}
\usepackage{rotating}

\begin{document}

\title{Testing several recent van der Waals density functionals for layered structures}
\author{Torbj\"orn Bj\"orkman}
 \affiliation{COMP Centre of Excellence, Department of Applied Physics, Aalto University School of Science, P.O. Box 11100, 00076 Aalto, Finland}
\email{torbjorn.bjorkman@aalto.fi}
\date{\today}

\begin{abstract}
Six recently developed exchange functionals for pairing with different two versions of van der Waals density functionals (vdW-DF) are tested for weakly bonded solids. The test, using 26 layered weakly bonded compounds, benchmarks the lattice constants against experimental data and the interlayer binding energies against reference data from the random-phase approximation (RPA). The investigated functionals tend to give interlayer binding energies higher than the RPA benchmark, and the overall performance for lattice constants is good. The exchange functionals optB86b and cx13 paired with the original vdW-DF and the B86R functional paired with vdW-DF2 are found to give particularly good results for equilibrium geometries.
\end{abstract}


\maketitle

\section{Introduction}

The van der Waals density functional (vdW-DF) method of Dion et al.\cite{Dion2004} is emerging as one of the most successful methods for including attractive long-range dispersion forces in practical density functional theory calculations. The scheme describes the long-range interaction by letting the density at all points in space explicitly interact with all other points through a kernel function that locally models the dielectric response at each point by means of a plasmon-pole approximation\cite{Dion2004,langreth2005,Lee2010,Berland2014a,Berland2014b}. This is then combined with a generalized-gradient approximation (GGA) for the exchange energy and the local density approximation (LDA) for the local part of the correlation, a construction that ensures that the functional reverts to LDA in the homogenous electron gas limit. The original method was followed by a revised version by Lee et al.\cite{Lee2010} (vdW-DF2) of the interaction kernel that modifies the plasmon-pole approximation to a form better suited for molecules. 

The performance of the functional depends critically on the choice of the GGA exchange functional to be paired with the non-local correlation. The original choice was the revPBE functional\cite{revPBE}, selected on the basis that it gives very little binding from exchange, but which also consistently overestimates the bond lengths of van der Waals bonded systems. A new exchange functional was developed by Murray et al.\cite{Murray2009} for pairing with the vdW-DF2 kernel, a reparametrization of the PW86 functional\cite{pw86} labelled PW86R, constructed to accurately represent the long-range behaviour of the exact exchange. The vdW-DF2 kernel with PW86R exchange gives an improvement of bond lengths, but still significantly overestimates them in many cases\cite{Bjorkman2012b}. The first attempt to construct a functional directly for the vdW-DF non-local correlation was done by Cooper\cite{Cooper2010}, who designed a functional, C09, that retains the long range behaviour of revPBE while decreasing the short-range repulsion. A different set of functionals was designed by Klime{\v{s}}, Michaelides and co-workers by optimization of different functional forms to match interaction energies for molecules\cite{Klimes2010,Klimes2011}, named by prefixing the parent functionals by ''opt''. In an attempt to construct an exchange pairing based on considerations similar to those that lead to the non-local vdW-DF kernel, Berland and Hyldgaard constructed the cx13 functional\footnote{The name of cx13 is intended to denote ''consistent exchange'' and 13 is the year of its construction. It is so not an acronym for the authors, hence the lower case letters.}. This was based on matching the exchange gradient expansion to the internal functional that generates the plasmon-pole approximation inherent in the vdW-DF kernel and then using insights gained from extensive benchmarking\cite{Berland2013} to combine this with PW86R to form a practically working scheme\cite{Berland2014a}. Even more recently, Hamada constructed a GGA exchange pairing for vdW-DF2 based on B86b by somewhat similar design principles, based on considerations of suitable forms for the small and large reduced density gradient regions (here designated revB86b-DF2)\cite{Hamada2014}. These functionals have been demonstrated to yield good results for small molecules\cite{Cooper2010,Berland2013,Klimes2010,Klimes2011,Hamada2014}, solids\cite{Berland2013,Klimes2010,Klimes2011,Hamada2014}, some layered systems\cite{Cooper2010,Graziano2012,Berland2014a,Ding2012,Mirhosseini2014,Peelaers2014} and the optB88 and optB86b functionals have been successfully applied to adsorption problems\cite{Mittendorfer2011,Luder2014a}. In relation to layered compounds can also be mentioned that promising results for graphite has been reported with the rVV10 functional\cite{rVV10} as well as with a pairing of the C09 functional with the vdW-DF2 kernel\cite{Hamada2010,Mapasha2012}.

The present author and co-workers have in two previous papers investigated vdW interaction in layered, weakly bonded systems\cite{Bjorkman2012a,Bjorkman2012b} in terms of their equilibrium geometries, compared with experimental structures, and the interlayer binding energy, compared to RPA. This data was also used to construct two revised versions of VV10 non-local correlation functional\cite{vv10}, AM05-VV10sol and PW86r-VV10sol\cite{Bjorkman2012c}, of which the former was shown to perform well both for equilibrium geometries and interlayer binding energies in layered materials. These studies did not include the later developments discussed above, and here an attempt is made to fill this gap in the knowledge of the performance of these methods for layered vdW solids. The present work investigates C09, cx13, revB86b-DF2 and the three ''opt'' functionals optPBE, optB88 and optB86b by the same benchmark suite used earlier for the development of the VV10sol functionals. The interlayer binding energy is compared with RPA values and the equilibrium geometries are compared to experimental values.

\section{Methods}

All calculations were carried out using the projector-augmented wave (PAW) method as implemented in the {\sc vasp} software package\cite{vasp1,vasp2,vasppaw} with an in-house implementation of vdW-DF using an adaptive real-space grid technique\cite{gulans2009}. The cx13 functional was implemented in {\sc vasp} for the present work\footnote{A patch for inclusion of the cx13 functional in {\sc VASP} version 5.3 is available from the author upon request.}, while all other functionals correspond to specific choices of parameters of previously implemented functionals. The basis set plane wave cutoff was set to 1.5 times the default given in the PAW library and Brillouin zone integrations were carried out on a k-space mesh with spacing of 0.2\AA$^{-1}$ and a gaussian broadening of 0.1 eV. As the vdW-DF implementation does not include stress tensor calculation, equilibrium geometries were determined by minimizing the energy with respect to the crystallographic parameters using a downhill simplex method, while fully relaxing the internal coordinates in each step. The internal coordinates were relaxed until residual forces were smaller than 0.01eV/\AA{} and the downhill simplex method was terminating when the relative change in lattice constants were smaller than 0.001 and the relative change in energy was smaller than 0.1. 

The binding energy was determined by calculation of the total energy as function of interlayer separation of the layers with the intraplanar geometry fixed at the experimental structure, as was previously done in the reference RPA calculations\cite{Bjorkman2012a}. The choice of the RPA as a benchmark for the binding energies is a matter of computational constraints; it is still the only higher order method that is feasible to apply to a set of solids as large as the one used here. However, from basic theoretical considerations the RPA is expected to describe the vdW part of the interaction\cite{Dobson2012}. There are presently very few binding energy calculations for layered systems available going beyond RPA. To the best of my knowledge, the only cases are quantum Monte Carlo results for graphite\cite{Spanu2009} (stronger binding than RPA) and TD-DFT results from Olsen and Thygesen for bilayer graphene\cite{Olsen2013}(weaker binding than RPA) and graphene on Ni\cite{Olsen2014} (chemisorption as strong as RPA and physisorption weaker than RPA). The scarcity of data and the very special electronic structure of graphene planes make these of limited value for establishing the accuracy of RPA for layered systems, but at least they are close to the RPA in value and do not show any systematic errors.

Since the shape of the interlayer binding energy curve is rather asymmetric, the lowest vibrational energy eigenvalues are sufficiently high up from the bottom of the potential well for anharmonic effects to be apparent already in the ground state. The results should therefore be corrected for the resulting zero-point anharmonic expansion (ZPAE) In order to get an accurate comparison to experimental values of the out-of-plane $c$ lattice constants. Following Bauer and Wu\cite{Bauer1956}, we may estimate the effect of zero point vibrations by fitting the binding energy curve near its minimum to a Morse potential, 
\begin{equation}
V(d) = D \left(1-e^{-a(d-d_0)}\right)^2 + C,
\end{equation}
which has known analytic solutions for the vibrational spectrum.  The additional constant $C$ is a convenience to get an accurate fit near the minimum and it should be noted that the normal interpretation of the parameter $D$ as the binding energy of the system is not correct when fitting the binding energy curve in this way. This is due to the asymptotic behavior of the Morse potential, which is exponential rather than polynomial, as would be correct for van der Waals bonded systems. We then use Equations (5) and (9) of Bauer and Wu to find $\Delta d$, the average displacement from equilibrium of a low lying Morse oscillator energy level $n$,
\begin{equation}
\Delta d_n = \frac{3\hbar\omega}{4 a D}\left(\frac{1}{2}+n\right).
\end{equation}
For the Morse oscillator, $\omega=a\sqrt{\frac{2D}{\mu}}$, with $\mu$ being the reduced mass of the system, which for the ground state gives of a system of layers of similar masses yields,
\begin{equation}\label{deltad}
\Delta d_n = \frac{0.068576}{\sqrt{D\mu}}\left(\frac{1}{2}+n\right) \text{\AA},
\end{equation}
where $\mu$ is measured in atomic mass units per unit cell and layer, and $D$ is measured in eV per layer. Similarly, the correction to the binding energy for the lowest vibrational levels is given by the normal harmonic oscillator expression\cite{Bauer1956} with a suitable substitutions of Morse potential parameters made for $\omega$,
\begin{equation}\label{vibrenergy}
E_n = \hbar\omega \left( \frac{1}{2}+n\right)= 0.091435\cdot a \sqrt{\frac{D}{\mu}} \left(\frac{1}{2}+n\right)\text{eV,}
\end{equation}
where $a$ is measured in \AA$^{-1}$.
The method of fitting the binding curve to a Morse potential and then applying equations \eqref{deltad} and \eqref{vibrenergy} should provide useful estimates lowest vibrational states for many other systems, such as adsorption on surfaces and layered heterostructures, as long as a reduction of the out-of-plane motion to a single degree of freedom can be done.

\section{Results and discussion}

\begin{table*}[htdp] 
\caption{Summary of mean relative errors (MRE) and mean absolute relative errors (MARE) for the investigated functionals. Errors in binding energy $<15$\% and errors in crystallographic parameters $<1$\% are shown in bold. Large errors in binding energy, $>25$\% in binding energy and $>4$\% in crystallographic parameters, are underlined. }
\label{summarytable}
\begin{center}
\begin{tabular}{l|cccc|cc|cc}
Functional & \multicolumn{4}{c}{$E_B$} & \multicolumn{2}{c}{$c$ axis} & \multicolumn{2}{c}{$a$ axis} \\
                  & ME (meV/\AA$^2$) & MAE (meV/\AA$^2$) & MRE (\%) & MARE (\%) & MRE (\%) & MARE (\%) & MRE (\%) & MARE (\%) \\\hline\hline
C09            & 8.9 & 8.9 & \underline{46} & \underline{46} & \textbf{-0.6} & 1.0 & -1.6 & 1.6 \\
cx13          & 4.4 & 4.4 & 22 & 22 & \textbf{-0.4} & \textbf{0.9} & \textbf{-0.6} & \textbf{0.9} \\
optPBE     & 0.8 & 2.3 & \textbf{6.3} & \textbf{12} & 1.3 & 1.4 & \underline{4.4} & \underline{4.4} \\
optB88     & 4.1 & 4.2 & 23 & 23 & 1.0 & 1.0 & 1.8 & 1.9 \\
optB86b   & 5.1 & 5.1 & \underline{27} & \underline{27} & \textbf{0.1} & \textbf{0.6} & \textbf{0.6} & 1.0 \\
revB86b-DF2 & 2.6 & 2.9 & \textbf{14} & 16 & \textbf{0.2} & \textbf{0.6} & \textbf{0.3} & \textbf{0.9} \\\hline
AM05-VV10sol & & & \textbf{5.2} & \textbf{11} & \textbf{-0.1} & 1.6 & -1.2 & 1.4 \\
vdW-DF & -4.1 & 4.4 & -18 & 20 & \underline{7.7} & \underline{7.7} & 2.5 & 2.5 \\
vdW-DF2 & -3.1 & 3.5 & \textbf{-13} & \textbf{15} & \underline{5.8} & \underline{5.8} & \underline{4.2} & \underline{4.2}
\end{tabular}
\end{center}
\end{table*}

\begin{figure}[htbp] 
   \centering
   \includegraphics[width=0.49\textwidth]{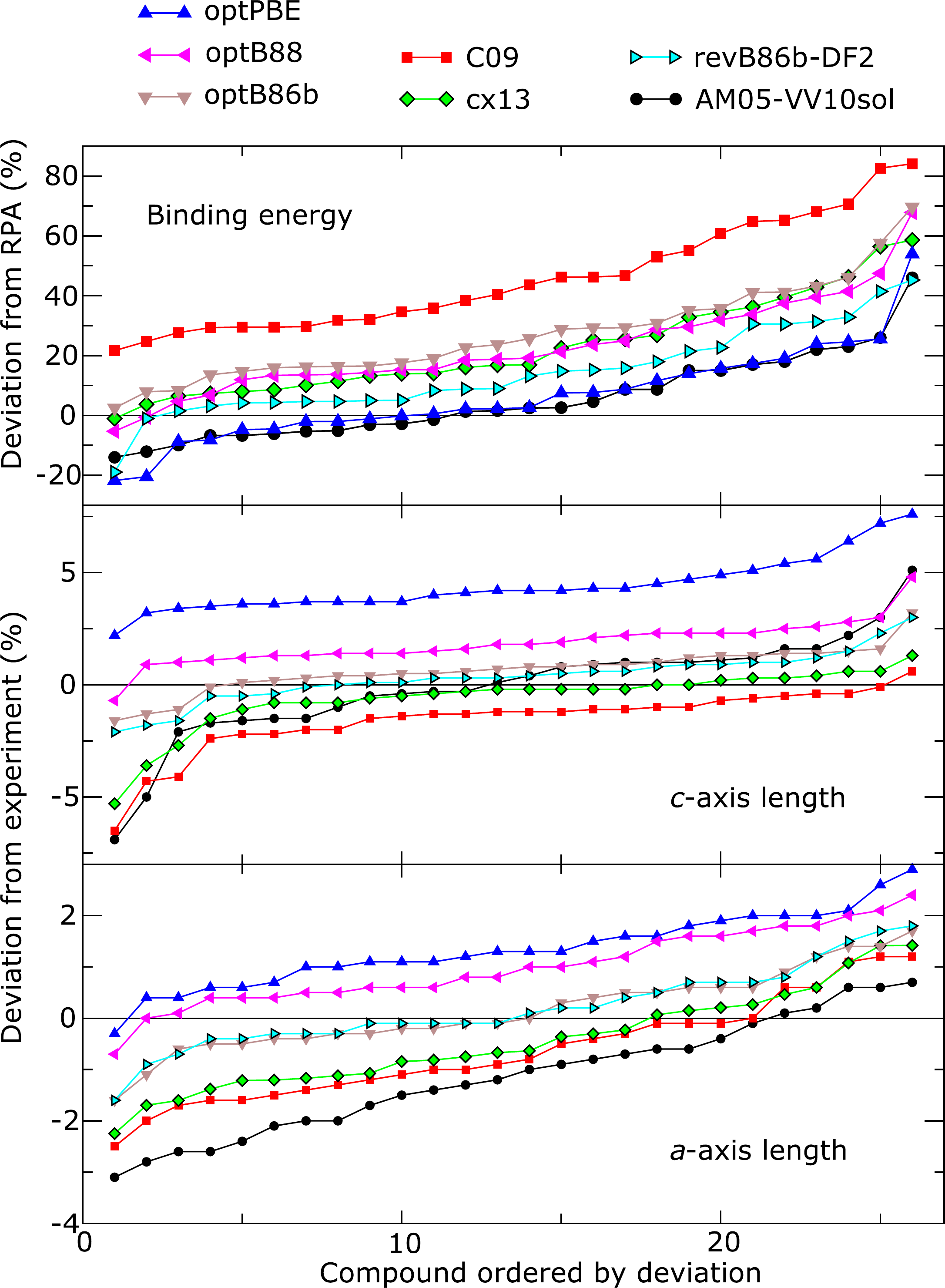} 
   \caption{Deviations from RPA binding energies and experimental lattice constants, sorted by the size of the deviation. Note that the index along the $x$-axes do not correspond to the same compound for the different functionals.}
   \label{errors}
\end{figure}

The results of all the investigated functionals are given in Tables \ref{energies}, \ref{c-constants} and \ref{a-constants}. For the $c$-axes lengths in Table \ref{c-constants}, the ZPAE have been individually calculated and added to the result using the same functional as was used in the relaxation. The errors of the different functionals are summarized in Table \ref{summarytable} in terms of the mean relative errors (MRE) and mean absolute relative errors (MARE) of the binding energies and crystallographic parameters. The mean errors and absolute mean errors compared to the RPA reference are also listed for the binding energies\footnote{Since the systems under consideration are not homogenous in shape, the deviations of the lattice constants cannot be directly compared since, unlike for the interlayer binding energies which are measured per layer and unit area, there is no sensible normalization common to all cases.}.
For reference is also included results for the previously investigated functionals vdW-DF\cite{Dion2004}, vdW-DF2\cite{Lee2010} and AM05-VV10sol\cite{Bjorkman2012c}. The smallest average deviations have been highlighted by boldface and the largest deviations by underlining. A graphical illustration of the deviations is also given in Figure \ref{errors}, where compounds have been sorted from the smallest to largest value of deviation. 

The convergence of these settings was tested by recalculating the revB86b-DF2 with a 15-20\% basis set cutoff and on a k-space mesh with a tighter spacing of 0.15\AA$^{-1}$. This induced fluctuations of the $c$ lattice constants at most by $\pm0.3$\%, which is a good measure of the uncertainty of the individual compounds. The convergence errors of the $a$ lattice constants are about one order of magnitude smaller, ~0.01\%, and for $E_B$ they were found to be $\pm 0.02$meV/\AA$^2$ or ~0.1\%. These errors showed no drift and do not alter the average values listed in Table \ref{summarytable}. We may also note the excellent agreement of the values for optPBE and optB88 functionals for graphite and BN compared with those of Graziano et al.\cite{Graziano2012}, using the same electronic structure code, but a different implementation of the vdW-DF framework. However, the agreement for cx13 with the calculations of Berland and Hyldgaard\cite{Berland2014a}, which were performed using the Quantum Espresso package\cite{quantumespresso}, is not as good, despite the fortran subroutine of the present implementation of the cx13 functional being verified to give the same results as that of Berland and Hyldgaard to machine precision. The most probable source of this difference appears to be the different pseudopotential libraries employed.

The functionals all give ZPAE of similar sizes for the different compounds. Indeed, within the accuracy of the present study, the correction could have been calculated with any one of the functionals and then applied instead to the experimental values. For reference, ZPAE corrected experimental values are given in parentheses in Table \ref{c-constants}, and zero point energy corrected RPA binding energies are given in Table \ref{energies}, with zero-point corrections taken from the revB86b-DF2 functional. The overall effect of ZPAE is to shift the average errors in $c$ axis length by approximately 0.1\% for all functionals, with effects being very small in all compounds except graphite and h-BN, where it is approximately 1\%, in agreement with a previous estimate of Graziano et al.\cite{Graziano2012}. The reason is obvious from the form of Equation \eqref{deltad}, which has the square root of the reduced mass in the denominator, thus greatly reducing the effect for all but the lightest elements.

Table \ref{summarytable} and Figure \ref{errors} gives a rather clear gradation of the functionals' performance for the different properties. The C09 functional is clearly overbinding, with binding energies far above the RPA values and a systematic underestimate of the lattice constants. The optPBE functional comes closest to the RPA binding energies, but has rather too large lattice constants, with average in-plane $a$ lattice constants being very much too large. The optB88 functional, which was shown to very closely reproduce RPA for the binding behaviour of graphene on Ni\cite{Mittendorfer2011}, is here found to consistently give much higher binding energies, yet still produces too large lattice constants. The three functionals optB86b, cx13 and revB86b-DF2 all perform very well for the investigated equilibrium geometries, with binding energies higher than the RPA benchmark, in the case of optB86b as high as 27\% higher. 

Summarizing the results into practical recommendations, the rather strong overbinding of C09 and the large overestimate of in-plane lattice constants of optPBE speak against the use of these, while any of the three functionals optB86b, cx13 or revB86b-DF2 could be expected to perform well in layered systems, primarily based on the excellent agreement with the experimental geometries.. They all have higher interlayer binding energies than the reference RPA calculations, but it should also be noted that RPA is not an exact method, merely the best available benchmark. Furthermore, correct equilibrium geometries is expected to be of greater importance than binding energies in most typical applications.

\begin{table*}[htdp]
\caption{Binding energies for 26 weakly bonded layered compounds with RPA reference energies are from Ref. \cite{Bjorkman2012a}. Energies are given in  meV/\AA$^2$ and the mean relative error (MRE) and the mean of the absolute error (MARE) are given in percent. For the reference energy $E_B^{RPA}$, the binding energy including the vibrational zero-point energy (calculated by revB86b-DF2) is given in parentheses.}
\begin{center}
\begin{tabular}{l|c|cc|cc|cc|cc|cc|cc}
    \multicolumn{2}{c}{} & \multicolumn{2}{c}{C09\cite{Cooper2010}} & \multicolumn{2}{c}{cx13\cite{Berland2014a}} & \multicolumn{2}{c}{optPBE\cite{Klimes2010}} & \multicolumn{2}{c}{optB88\cite{Klimes2010}} & \multicolumn{2}{c}{optB86b\cite{Klimes2011}} & \multicolumn{2}{c}{revB86b\cite{Hamada2014}}\\
   Compound  & $E_B^{RPA}$ & $E_B$ &  RE  & $E_B$ &  RE &  $E_B$ &  RE  & $E_B$ &  RE  & $E_B$ &  RE & $E_B$ &  RE  \\\hline
    Graphite &  18.3 (17.2) & 26.9 &  47  & 23.0 &   25 & 22.7 &    24 & 25.1 &   38 & 24.8 & 35  & 21.0 &   15 \\
      BN     &  14.4 (13.4) & 26.7 &  84  & 22.7 &   56 & 22.3 &    54 & 24.3 &   68 & 24.5 & 69  & 20.5 &   41 \\  
    HfS$_2$  &  16.1 (15.9) & 22.7 &  40  & 18.7 &   16 & 18.9 &    17 & 20.9 &   30 & 20.8 & 29  & 18.3 &   13 \\
   HfSe$_2$  &  17.0 (16.9) & 23.2 &  36  & 19.3 &   13 & 18.3 &   7.5 & 20.7 &   21 & 20.9 & 22  & 18.6 &    9 \\
   HfTe$_2$  &  18.6 (18.5) & 27.3 &  46  & 23.4 &   25 & 19.0 &   2.2 & 22.2 &   19 & 23.4 & 25  & 21.5 &   15 \\
    MoS$_2$  &  20.5 (20.2) & 26.6 &  30  & 22.0 &  7.3 & 20.9 &   2.1 & 23.6 &   15 & 23.9 & 16  & 21.6 &    5 \\
   MoSe$_2$  &  19.6 (19.4) & 25.9 &  32  & 21.6 &   10 & 19.4 &  -1.1 & 22.2 &   14 & 22.8 & 16  & 20.6 &    5 \\
   MoTe$_2$  &  20.8 (20.6) & 27.5 &  32  & 23.7 &   14 & 19.0 &  -8.1 & 22.2 &  6.8 & 23.6 & 13  & 21.7 &    4 \\
   NbSe$_2$  &  19.5 (19.3) & 32.3 &  65  & 27.3 &   39 & 22.2 &    14 & 26.1 &   34 & 27.6 & 41  & 25.5 &   31 \\
   NbTe$_2$  &  23.0 (22.9) & 35.7 &  55  & 31.4 &   36 & 22.9 &  -0.1 & 27.3 &   19 & 30.1 & 30  & 28.2 &   23 \\
     PbO     &  20.2 (20.1) & 24.6 &  22  & 20.0 & -1.0 & 16.1 &   -20 & 20.1 & -0.6 & 20.7 & 2.5 & 16.4 &  -19 \\
   PdTe$_2$  &  40.1 (39.9) & 52.0 &  29  & 47.0 &   17 & 31.4 &   -22 & 38.0 & -5.2 & 43.3 & 7.9 & 40.8 &    2 \\
    PtS$_2$  &  20.5 (20.4) & 29.5 &  44  & 23.4 &   14 & 19.5 &  -4.7 & 23.3 &   14 & 24.1 & 17  & 21.5 &    5 \\
   PtSe$_2$  &  19.0 (18.9) & 31.5 &  65  & 25.6 &   35 & 18.1 &  -4.5 & 22.5 &   18 & 24.6 & 29  & 22.0 &   16 \\
    TaS$_2$  &  17.6 (17.4) & 28.4 &  61  & 23.5 &   33 & 22.0 &    25 & 25.0 &   41 & 25.3 & 43  & 23.1 &   31 \\
   TaSe$_2$  &  19.4 (19.2) & 28.4 &  46  & 23.8 &   23 & 21.1 &   8.7 & 24.2 &   25 & 25.0 & 28  & 22.9 &   18 \\
    TiS$_2$  &  18.8 (18.5) & 28.8 &  53  & 23.9 &   27 & 21.8 &    16 & 24.9 &   32 & 25.5 & 35  & 22.9 &   21 \\
   TiSe$_2$  &  17.3 (17.1) & 29.6 &  71  & 24.8 &   43 & 20.7 &    19 & 24.2 &   40 & 25.4 & 46  & 23.1 &   33 \\
   TiTe$_2$  &  19.7 (19.5) & 33.2 &  68  & 28.9 &   46 & 21.2 &   7.6 & 25.4 &   29 & 27.9 & 41  & 26.0 &   31 \\
     VS$_2$  &  25.6 (25.3) & 31.9 &  25  & 26.6 &  3.7 & 23.3 &  -8.7 & 26.8 &  4.7 & 27.7 & 8.3 & 25.3 &   -1 \\
    VSe$_2$  &  22.2 (22.0) & 30.7 &  38  & 26.0 &   17 & 21.8 &  -2.0 & 25.2 &   13 & 26.5 & 19  & 24.3 &    9 \\
     WS$_2$  &  20.2 (20.0) & 25.8 &  28  & 21.6 &  6.4 & 20.7 &   2.6 & 23.3 &   15 & 23.5 & 16  & 21.2 &    5 \\
    WSe$_2$  &  19.9 (19.7) & 25.9 &  30  & 21.7 &  9.5 & 19.5 &  -2.0 & 22.3 &   12 & 22.9 & 14  & 20.8 &    4 \\
    ZrS$_2$  &  16.9 (16.7) & 22.8 &  35  & 18.9 &   11 & 18.9 &    12 & 21.0 &   24 & 20.9 & 23  & 18.4 &    8 \\
   ZrSe$_2$  &  18.5 (18.3) & 24.0 &  30  & 20.0 &  8.0 & 18.6 &   0.5 & 21.1 &   14 & 21.4 & 15  & 19.1 &    3 \\
   ZrTe$_2$  &  16.3 (16.1) & 29.8 &  83  & 25.9 &   59 & 20.5 &    26 & 24.0 &   47 & 25.7 & 57  & 23.7 &   45 \\\hline
     MRE     &        &      &  46  &      &   22 &      &   6.3 &      &   23 &      & 27  &      &   14 \\
    MARE     &        &      &  46  &      &   22 &      &    12 &      &   23 &      & 27  &      &   16
\end{tabular}
\end{center}
\label{energies}
\end{table*}%

\begin{table*}[htdp]
\caption{Out-of-plane lattice constants, $c$ in \AA{} and the mean relative error (MRE) and the mean of the absolute error (MARE) given in percent. The appropriate ZPAE corrections for each functional have been applied to the calculated values and for the experimental data, the values corrected for ZPAE (as calculated by revB86b-DF2) are given in parentheses. }
\begin{center}
\begin{tabular}{l|c|cc|cc|cc|cc|cc|cc}
    \multicolumn{2}{c}{} & \multicolumn{2}{c}{C09} & \multicolumn{2}{c}{cx13} & \multicolumn{2}{c}{optPBE} & \multicolumn{2}{c}{optB88} & \multicolumn{2}{c}{optB86b} & \multicolumn{2}{c}{revB86b} \\
   Compound  & $c^{exp}$ & $c$ &  RE  & $c$ &  RE &  $c$ &  RE  & $c$ &  RE  & $c$ &  RE  & $c$ &  RE  \\\hline
    Graphite &  6.696 (6.635) & 6.54  & -2.4 & 6.65  & -0.8 & 6.98  & 4.3 & 6.76  & 1.0  & 6.72  & 0.4  & 6.72  & -0.5 \\
      BN     &  6.690 (6.635) & 6.42  & -4.1 & 6.51  & -2.7 & 6.84  & 2.2 & 6.64  & -0.7 & 6.58  & -1.6 & 6.60  & -2.1 \\  
    HfS$_2$  &  5.837 (5.830) & 5.77  & -1.2 & 5.84  & 0.0  & 6.05  & 3.6 & 5.91  & 1.2  & 5.87  & 0.5  & 5.85  & 0.1  \\
   HfSe$_2$  &  6.159 (6.153) & 6.09  & -1.2 & 6.15  & -0.2 & 6.39  & 3.7 & 6.25  & 1.4  & 6.22  & 0.9  & 6.19  & 0.3  \\
   HfTe$_2$  &  6.650 (6.645) & 6.58  & -1.0 & 6.63  & -0.2 & 6.97  & 4.9 & 6.80  & 2.3  & 6.69  & 0.5  & 6.70  & 0.6  \\
    MoS$_2$  & 12.302 (12.283) & 12.18 & -1.0 & 12.34 & 0.3  & 12.81 & 4.1 & 12.53 & 1.8  & 12.41 & 0.9  & 12.38 & 0.5  \\
   MoSe$_2$  & 12.927 (12.912) & 12.86 & -0.5 & 12.98 & 0.4  & 13.51 & 4.5 & 13.22 & 2.3  & 13.10 & 1.4  & 13.06 & 1.0  \\
   MoTe$_2$  & 13.973 (13.961) & 13.87 & -0.7 & 13.95 & -0.2 & 14.56 & 4.2 & 14.34 & 2.6  & 14.11 & 1.0  & 14.12 & 1.0  \\
   NbSe$_2$  & 12.547 (12.534) & 12.29 & -2.0 & 12.36 & -1.5 & 13.05 & 4.0 & 12.74 & 1.6  & 12.55 & 0.1  & 12.49 & -0.5 \\
   NbTe$_2$  &  6.610 (6.606) & 6.65  &  0.6 & 6.69  & 1.3  & 7.08  & 7.2 & 6.92  & 4.8  & 6.82  & 3.2  & 6.81  & 3.0  \\
     PbO     &  4.995 (4.990) & 4.88  & -2.2 & 4.96  & -0.6 & 5.28  & 5.6 & 5.10  & 2.2  & 5.07  & 1.4  & 5.12  & 2.3  \\
   PdTe$_2$  &  5.113 (5.110) & 5.09  & -0.4 & 5.12  & 0.2  & 5.27  & 3.2 & 5.25  & 2.8  & 5.15  & 0.8  & 5.16  & 0.9  \\
    PtS$_2$  &  5.043 (5.036) & 4.72  & -6.5 & 4.78  & -5.3 & 5.37  & 6.4 & 5.12  & 1.4  & 4.99  & -1.1 & 4.96  & -1.8 \\
   PtSe$_2$  &  5.081 (5.074) & 4.86  & -4.3 & 4.90  & -3.6 & 5.46  & 7.6 & 5.15  & 1.3  & 5.02  & -1.3 & 5.01  & -1.6 \\
    TaS$_2$  &  5.897 (5.890) & 5.83  & -1.2 & 5.89  & -0.2 & 6.15  & 4.2 & 6.01  & 1.9  & 5.97  & 1.2  & 5.94  & 0.6  \\
   TaSe$_2$  &  6.272 (6.267) & 6.20  & -1.1 & 6.25  & -0.4 & 6.51  & 3.7 & 6.37  & 1.5  & 6.32  & 0.7  & 6.29  & 0.1  \\
    TiS$_2$  &  5.705 (5.696) & 5.58  & -2.2 & 5.66  & -0.8 & 5.91  & 3.6 & 5.77  & 1.1  & 5.73  & 0.4  & 5.69  & -0.4 \\
   TiSe$_2$  &  6.004 (5.996) & 5.89  & -2.0 & 5.94  & -1.1 & 6.26  & 4.2 & 6.09  & 1.4  & 6.00  & -0.1 & 6.01  & -0.1 \\
   TiTe$_2$  &  6.498 (6.493) & 6.40  & -1.4 & 6.46  & -0.5 & 6.74  & 3.7 & 6.64  & 2.1  & 6.51  & 0.3  & 6.52  & 0.3  \\
     VS$_2$  &  5.755 (5.748) & 5.69  & -1.1 & 5.74  & -0.3 & 6.05  & 5.1 & 5.93  & 3.0  & 5.83  & 1.3  & 5.81  & 0.8  \\
    VSe$_2$  &  6.107 (6.101) & 6.07  & -0.6 & 6.13  & 0.3  & 6.44  & 5.4 & 6.25  & 2.3  & 6.20  & 1.5  & 6.19  & 1.2  \\
     WS$_2$  & 12.323 (12.308) & 12.27 & -0.4 & 12.40 & 0.6  & 12.86 & 4.3 & 12.60 & 2.3  & 12.49 & 1.3  & 12.46 & 0.9  \\
    WSe$_2$  & 12.960 (12.947) & 12.95 & -0.1 & 13.04 & 0.6  & 13.56 & 4.7 & 13.28 & 2.5  & 13.16 & 1.6  & 13.16 & 1.5  \\
    ZrS$_2$  &  5.813 (5.805) & 5.73  & -1.5 & 5.80  & -0.2 & 6.02  & 3.5 & 5.87  & 0.9  & 5.85  & 0.6  & 5.84  & 0.3  \\
   ZrSe$_2$  &  6.128 (6.122) & 6.05  & -1.3 & 6.13  & 0.0  & 6.36  & 3.7 & 6.24  & 1.8  & 6.18  & 0.8  & 6.16  & 0.4  \\
   ZrTe$_2$  &  6.660 (6.655) & 6.57  & -1.3 & 6.60  & -0.8 & 6.89  & 3.4 & 6.75  & 1.3  & 6.67  & 0.2  & 6.66  & 0.0  \\\hline
     MRE     &        &       & -1.6 &       & -0.6 &       & 4.4 &       &  1.8 &        &  0.6 &       &  0.3 \\
    MARE     &        &       &  1.6 &       &  0.9 &       & 4.4 &       &  1.9 &        &  1.0 &       &  0.9
\end{tabular}
\end{center}
\label{c-constants}
\end{table*}%

\begin{table*}[htdp]
\caption{In-plane lattice constants, $a$ in \AA{} and the mean relative error (MRE) and the mean of the absolute error (MARE) given in percent.}
\begin{center}
\begin{tabular}{l|c|cc|cc|cc|cc|cc|cc}
    \multicolumn{2}{c}{} & \multicolumn{2}{c}{C09} & \multicolumn{2}{c}{cx13} & \multicolumn{2}{c}{optPBE} & \multicolumn{2}{c}{optB88} & \multicolumn{2}{c}{optB86b} & \multicolumn{2}{c}{revB86b} \\
   Compound  & $a^{exp}$ & $a$ &  RE  & $a$ &  RE &  $a$ &  RE  & $a$ &  RE  & $a$ &  RE  & $a$ & RE\\\hline
    Graphite & 2.456 & 2.47 &  0.6 & 2.47 &  0.4 & 2.47 &  0.6 & 2.47 &  0.6 & 2.47 &  0.6 & 2.47 &  0.4 \\
      BN     & 2.510 & 2.51 &  0.0 & 2.52 &  0.2 & 2.52 &  0.4 & 2.52 &  0.4 & 2.52 &  0.4 & 2.52 &  0.2 \\  
    HfS$_2$  & 3.635 & 3.59 & -1.2 & 3.60 & -0.8 & 3.66 &  0.7 & 3.65 &  0.4 & 3.62 & -0.4 & 3.62 & -0.3 \\
   HfSe$_2$  & 3.748 & 3.71 & -1.0 & 3.72 & -1.2 & 3.79 &  1.1 & 3.77 &  0.6 & 3.74 & -0.2 & 3.74 & -0.1 \\
   HfTe$_2$  & 3.957 & 3.89 & -1.7 & 3.90 & -1.1 & 4.01 &  1.3 & 3.99 &  0.8 & 3.94 & -0.4 & 3.94 & -0.4 \\
    MoS$_2$  & 3.162 & 3.15 & -0.4 & 3.15 & -0.3 & 3.20 &  1.2 & 3.20 &  1.2 & 3.17 &  0.3 & 3.17 &  0.2 \\
   MoSe$_2$  & 3.289 & 3.28 & -0.3 & 3.29 & -0.3 & 3.35 &  1.9 & 3.34 &  1.6 & 3.31 &  0.6 & 3.30 &  0.5 \\
   MoTe$_2$  & 3.518 & 3.50 & -0.5 & 3.51 & -0.2 & 3.59 &  2.0 & 3.58 &  1.8 & 3.54 &  0.6 & 3.54 &  0.7 \\
   NbSe$_2$  & 3.442 & 3.44 & -0.1 & 3.44 &  0.0 & 3.51 &  2.0 & 3.50 &  1.7 & 3.46 &  0.5 & 3.47 &  0.7 \\
   NbTe$_2$  & 3.680 & 3.65 & -0.8 & 3.66 & -0.6 & 3.74 &  1.6 & 3.72 &  1.1 & 3.68 &  0.0 & 3.68 &  0.1 \\
     PbO     & 3.964 & 4.01 &  1.2 & 4.02 &  1.4 & 4.08 &  2.9 & 4.06 &  2.4 & 4.03 &  1.7 & 4.03 &  1.8 \\
   PdTe$_2$  & 4.024 & 4.05 &  0.6 & 4.06 &  0.6 & 4.13 &  2.6 & 4.11 &  2.1 & 4.08 &  1.4 & 4.08 &  1.5 \\
    PtS$_2$  & 3.542 & 3.58 &  1.1 & 3.58 &  1.0 & 3.60 &  1.6 & 3.60 &  1.6 & 3.59 &  1.4 & 3.59 &  1.2 \\
   PtSe$_2$  & 3.727 & 3.77 &  1.2 & 3.78 &  1.4 & 3.80 &  2.0 & 3.80 &  2.0 & 3.77 &  1.2 & 3.79 &  1.7 \\
    TaS$_2$  & 3.364 & 3.32 & -1.3 & 3.33 & -1.0 & 3.40 &  1.1 & 3.39 &  0.8 & 3.36 & -0.1 & 3.36 & -0.1 \\
   TaSe$_2$  & 3.476 & 3.44 & -1.0 & 3.45 & -0.7 & 3.52 &  1.3 & 3.51 &  1.0 & 3.47 & -0.2 & 3.47 & -0.1 \\
    TiS$_2$  & 3.409 & 3.36 & -1.4 & 3.36 & -1.2 & 3.43 &  0.6 & 3.41 &  0.0 & 3.39 & -0.6 & 3.39 & -0.7 \\
   TiSe$_2$  & 3.536 & 3.48 & -1.6 & 3.50 & -1.1 & 3.57 &  1.0 & 3.55 &  0.4 & 3.52 & -0.5 & 3.52 & -0.3 \\
   TiTe$_2$  & 3.777 & 3.72 & -1.5 & 3.73 & -1.3 & 3.82 &  1.1 & 3.80 &  0.6 & 3.76 & -0.5 & 3.76 & -0.4 \\
     VS$_2$  & 3.221 & 3.14 & -2.5 & 3.15 & -2.2 & 3.21 & -0.3 & 3.20 & -0.7 & 3.17 & -1.6 & 3.17 & -1.6 \\
    VSe$_2$  & 3.358 & 3.29 & -2.0 & 3.30 & -1.6 & 3.37 &  0.4 & 3.36 &  0.1 & 3.32 & -1.1 & 3.33 & -0.9 \\
     WS$_2$  & 3.153 & 3.15 & -0.1 & 3.16 &  0.2 & 3.21 &  1.8 & 3.20 &  1.5 & 3.17 &  0.5 & 3.17 &  0.7 \\
    WSe$_2$  & 3.282 & 3.28 & -0.1 & 3.29 &  0.1 & 3.35 &  2.1 & 3.34 &  1.8 & 3.31 &  0.9 & 3.31 &  0.8 \\
    ZrS$_2$  & 3.662 & 3.63 & -0.9 & 3.64 & -0.6 & 3.70 &  1.0 & 3.68 &  0.5 & 3.66 & -0.1 & 3.66 & -0.1 \\
   ZrSe$_2$  & 3.770 & 3.73 & -1.1 & 3.74 & -0.8 & 3.82 &  1.3 & 3.79 &  0.5 & 3.76 & -0.3 & 3.77 & -0.1 \\
   ZrTe$_2$  & 3.952 & 3.89 & -1.6 & 3.90 & -1.6 & 4.01 &  1.5 & 3.99 &  1.0 & 3.94 & -0.3 & 3.94 & -0.3 \\\hline
     MRE     &       &      & -0.6 &      & -0.4 &      &  1.3 &      &  1.0 &      &  0.1 &      &  0.2 \\
    MARE     &       &      &  1.0 &      &  0.9 &      &  1.4 &      &  1.0 &      &  0.6 &      &  0.6
\end{tabular}
\end{center}
\label{a-constants}
\end{table*}%

\section{Acknowledgements}
Helpful discussions with Kristian Berland concerning the implementation of the cx13 exchange functional are gratefully acknowledged. The research was supported by the Academy of Finland through grant number 263416 and the COMP centre of excellence. Computational resources supplied by the Finnish IT Center for Science (CSC).

%

\end{document}